\documentstyle[12pt]{article}  
\begin{document}
\pagestyle{plain}\textheight=21truecm\textwidth=13truecm
\begin{center}
\begin{bf}
THE FORMATION AND ROLE OF VORTICES IN PROTOPLANETARY DISKS 
\end{bf}  
\end{center}
\vspace{1.cm}  
\begin{center}
Patrick Godon \& Mario Livio \\
\vspace{5.cm}  
Space Telescope Science Institute \\
3700 San Martin Drive \\ 
Baltimore, MD 21218 \\
\vspace{0.5cm}  
e-mail: godon@stsci.edu, mlivio@stsci.edu  \\ 
\end{center}
\newpage

\baselineskip 15pt plus 2pt
\begin{center}
\bf{Abstract}
\end{center}

We carry out a two-dimensional, compressible, simulation 
of a disk, including dust particles, 
to study the formation and role of vortices in protoplanetary disks.  
We find that anticyclonic vortices can form out of an initial random perturbation
of the vorticity field. Vortices have a typical decay time of the order of
50 orbital periods (for a viscosity parameter 
$\alpha = 10^{-4}$ and a disk aspect ratio of $H/r = 0.15$). 
If vorticity is continuously generated at a constant rate in the flow 
(e.g. by convection), then a large vortex can form and be sustained 
(due to the merger of vortices). 

We find that dust concentrates in the cores of vortices 
within a few orbital periods, when  
the drag parameter is of the order of the orbital frequency. 
Also, the radial drift of the dust induces a significant increase 
in the surface density of dust particles in the inner region of the disk.
Thus, vortices may represent the preferred location for planetesimal formation
in protoplanetary disks.  

We show that it is very difficult for vortex 
mergers to sustain a relatively coherent
outward flux of angular momentum. \\  

Subject Headings: accretion, accretion disks - circumstellar matter - 
hydrodynamics - planetary systems - stars: formation -
stars: pre-main sequence

\newpage 
\section{Introduction}

The interest in vortices in disks has recently regained 
momentum due to the potential role they could play in planet formation. 
It is believed that the initial process of planet formation takes place
via a progressive aggregation and sticking of dust grains in the 
primordial protoplanetary disk. The grain aggregates begin to 
settle toward the mid-disk plane and grow to centimeter-size grains.  
This process is efficient in producing centimer-sized
grains, but it is cut off sharply thereafter due to small scale turbulent diffusion
in the nebula. These particles form a massive dusty layer in the disk mid-plane. A 
gravitational instability of the dusty, dense mid-plane layer, can be triggered
by seeds of the order of meters to form 10-100 km planetesimals
(the radial velocity dispersion induced by drag will further delay the onset
of the instability until the mean size is in the range 10-100 m). 
Therefore, there is a gap (of at least two orders of magnitude) 
between the maximal particles size reached by coagulation (~centimeters) and 
the minimal
size required for planetesimal formation (~meters). 
In order to remedy this problem, it has been suggested 
(e.g. Barge \& Sommeria 1995; Adams \& Watkins 1995; Tanga et al. 1996) 
that dust concentration in the cores of anticyclonic vortices 
may be rapidly (within a few orbits) enhanced by a significant factor.
As a result, the size of the particles needed to trigger the gravitational
instability can be reduced by a comparable factor (for a review of the problem
see e.g. Tanga et al. 1996). Assuming that the drift is negligible in the vortex,  
a gap in the size of a factor of a hundred may perhaps be bridged in this way.
The cores of anticyclonic vortices may therefore be 
the preferred regions for rapid planetesimal (and planet core) formation.   

Simplified numerical simulations of vortices in disks 
(e.g. Bracco et al. 1998, solving the vorticity equation; 
Nauta 1999, using a shallow water equation) 
have shown that anticyclonic vortices may be rather stable and can survive
in the flow for many orbits.  
More detailed calculations (Godon \& Livio 1999; assuming a two-dimensional, 
compressible, viscous, polytropic disk) have shown explicitly that   
the exponential decay time of the vortices is inversely proportional to  
the alpha viscosity parameter 
($\alpha_{SS}$; Shakura \& Sunyaev 1973). Godon \& Livio found that the 
decay time can be of the order order
of $10-100$ orbits in protoplanetary disks (where it is believed that 
$\alpha_{SS} \approx
10^{-4} - 10^{-3}$, and the ratio of the disk thickness to the radial
distance is $H/r \approx 0.05-0.20$; e.g. Bell et al. 1995). 
Vortices can therefore live sufficiently long to allow (in principle at least) 
for dust to concentrate in their cores.  The results of  
recent, highly simplified simulations (Bracco et al.1999, still solving  
the vorticity equation) in fact point in that direction. 
However, the value of $H/r$ is not defined in Bracco et al. (1998, 1999).
The elliptical
vortices obtained by these authors are not very elongated, implying 
an effective aspect ratio of $H/r = 
c_s/v_K \approx 1 $ ($c_s$ is the speed of sound and $v_K$ is the Keplerian
velocity in the disk). In addition, the vorticity equation  
does not take into account the compressibility of the flow, and the
vortex interaction range is essentially infinite. \\ 

Most importantly, however, from the point of view of vortices as potential planet
formation sites, {\it it is presently not clear how vortices form in the 
disk initially}. 
An analysis carried out by Lovelace et al. (1999) 
suggests that a non-linear, non-axisymmetric,  
Rossby wave instability could lead to the formation of vortices.
Vortices can also appear in a broader context, e.g.
in galactic disks (Fridman \& Khoruzhii, 1999). 
However, so far, no process has been shown unambiguously 
to create vortices in disks around young stellar objects.   
The only robust numerical result obtained to date is that coherent 
anticyclonic vortices can survive in a strongly sheared (two-dimensional)
Keplerian flow
(for about 50 orbits when $\alpha_{SS} = 10^{-4}$, 
Godon \& Livio 1999). \\  

In the present work we first investigate the {\it formation} of vortices in a 
compressible Keplerian disk. We then perform simulations intended to study
the {\it dynamics of dust} in the presence of vortices. Finally, we examine the 
possible role of vortices in {\it angular momentum transport}. 
The technical details of the modeling are given in \S 2. 
The results are presented in \S 3, and a discussion follows.

\section{Accretion Disks modeling} 

We solve the time-dependent, vertically averaged equations of
the disk (e.g. Pringle 1981) using a pseudospectral method (Godon 1997). 
The equations are solved in the plane of the disk using 
cylindrical coordinates $(r, \phi)$. 
We use an alpha prescription for the viscosity law (Shakura
\& Sunyaev 1973), and assume a polytropic relation for the pressure
$P=K \rho^{1+1/n}$,
where $K$ is the polytropic constant and $n$ is the polytropic index. 
Details of the modeling, including the full equations and the numerical
method can be found in Godon (1997) and in Godon \& Livio (1999). 
Models with different values of the physical and numerical parameters 
have been run (see also Godon \& Livio 1999).   
However, 
in all the models presented here, we chose $H/r=0.15$, $\alpha_{SS} = 10^{-4}$,
$n=2.5$, with an initial density profile $\rho \propto r^{-15/8}$, 
and an initial Keplerian angular velocity 
(standard disk model, Shakura \& Sunyaev 1973). A numerical resolution
of $128 \times 128$ collocation points has been used.  

\subsection{The Viscous Polytropic Disk} 

Since we are modeling a viscous polytropic disk, it is important to
stress that vorticity is dissipated and cannot be generated, for the 
following reason. 
The equation for the vorticity $\vec{\omega} = \vec{\nabla} \times \vec{v}$  
can be obtained by taking the curl of the Navier-Stokes equations 
(e.g. Tassoul 1978):   
\begin{equation}
\frac{D}{D t} \frac{\vec{\omega}}{\rho} = 
\frac{\vec{\omega}}{\rho}.\vec{\nabla} \vec{v} - 
\frac{1}{\rho} \vec{\nabla} \frac{1}{\rho} \times \vec{\nabla}P 
+ \vec{\nabla} \times \frac{1}{\rho} \vec{\nabla}.\vec{\tau},  
\end{equation} 
where 
\[ 
\frac{D}{D t} \frac{\vec{\omega}}{\rho} =  
\frac{\partial}{\partial t} \frac{\vec{\omega}}{\rho} + 
\vec{v}.\vec{\nabla} \frac{\vec{\omega}}{\rho}  ,~~~~~~~~~~~~~~~~~~~~~~~~~~~~~~~~~~~~~~~~~~~~~~~~~~~~~~~~~~(1')
\]
and $\vec{\tau}$ is the viscous stress tensor. The second term
on the RHS of equation (1) is a source term for the vorticity. This 
term is non-zero in a baroclinic flow [when
$P=P(\rho, T)$] and it vanishes in a barotropic flow [$P=P(\rho)$]. 
The last term on the RHS is the curl
of the viscous forces and it is responsible for the viscous dissipation
of the vorticity. Consequently, in an inviscid 
barotropic flow, the flux of vorticity across a material surface  
is a conserved quantity (this is known as Kelvin's circulation theorem).  
In the present work, the flow is polytropic $P=K \rho^{\gamma}$, and 
viscous, therefore vorticity is only dissipated.   

\subsection{Dust Modelling} 

In this work we also address the question of the concentration of
dust particles in the cores of anticylconic vortices. 
The equations for the dust "particles"
were simplified, by taking into account 
only the drag force exerted by the gas on the dust particles
(see e.g. discussion in Barge \& Sommeria 1995; Tanga et al. 1996).
We also assumed the radius $s$ of the dust particles to be smaller than
the mean free path $\lambda$ of the gas molecules (this is known as the 
Epstein regime, see e.g. Cuzzi, Dobrovolskis and Champney, 1993).  
Accordingly, the equations of motion of a dust particle located 
in the plane ($r, \phi$) are
given in the inertial frame of reference by      
\begin{equation} 
\frac{d^2r}{dt^2} = r \left( \frac{d \phi}{dt} \right)^2 
- \frac{GM}{r^2} - \gamma \left( \frac{dr}{dt} -v_r \right) , 
\end{equation} 
\begin{equation} 
r \frac{d^2 \phi}{dt^2} = -2 \frac{dr}{dt} \frac{d \phi}{dt}  
- \gamma \left( r \frac{d \phi}{dt} -v_{\phi} \right) , 
\end{equation} 
where $G$ is the gravitational constant, $M$ is the mass of the central
star, $v_r$ and  
$v_{\phi}$ are the radial and angular (respectively) components  
of the velocity of the flow and
$\gamma$ is the drag parameter. Sometimes it is convenient to write
$\gamma = \tau^{-1}$, where $\tau$ is the characteristic time for the 
dust to be dragged by the flow (or, in the frame co-moving with the flow, 
it is defined as the 'stopping' parameter). In the absence of drag 
($\gamma \rightarrow 0$), the equations represent the motion of  
particles in a Keplerian potential. 

\section{Results} 

\subsection{The Formation of vortices} 

In this subsection we simulate different mechanisms which are 
potentially capable 
of creating vortices in a thin, two-dimensional Keplerian disk
(representing a protoplanetary disk).  
We distinguish mainly between two types 
of processes: (i) vorticity is generated 
only initially (similar to simulations of decaying turbulence),    
(ii) vorticity is generated continuously (similar to  
simulations of driven turbulence). In the first case 
we assume that the flow is initially turbulent, while in the
second we propose two potential mechanisms for the generation
of vorticity: accretion
of clumps of gas (or "comets") onto the disk, and convection. 

\subsubsection{Initially Turbulent State} 

If the initial collapse of the protostellar cloud is turbulent, then one   
might suspect that the initial disk that forms could still have
some turbulence in it. It is common in the modeling of two-dimensional
turbulent flows (e.g. planetary atmospheres) to assume for the initial conditions
a random perturbation of the vorticity field 
(Bracco et al. 1998, 1999; Nauta 1999).   \\ 

We made a similar assumption for a standard disk model (Shakura 
\& Sunyaev 1973) with $H/r=0.15$ and $\alpha_{SS} = 10^{-4}$ 
(Figure 1). As the model
evolved, the anticyclonic vorticity perturbations formed coherent vortices
which merged together to form larger vortices, while the cyclonic vorticity 
perturbations were stretched and dissipated (Figure 2). 
The energy spectrum (Figure 3) was found to be 
fairly flat for small wave numbers, 
in agreement with previous simulations of two-dimensional compressible
turbulence (e.g. Farge \& Sadourni 1989; Godon 1998). The 
amplitude of the vortices decreased with time, as expected for a simulation
of decaying turbulence. The "turbulence" in this case
is not fed by the background
Keplerian flow, but rather only by the initial perturbation. \\ 

Our results are broadly 
consistent with those of Bracco et al. (1998). Namely, 
coherent anticyclonic vortices do form (from an initially perturbed vorticity
field) and are stable for many rotation
periods (they have exponential decay times of the order of $\approx 50$
orbits). However, a close examination of the results reveals further details.
In particular, we find that all the 
anticyclonic vortices are accompanied by cyclonic vorticity stripes, which 
form a partial shielding of the vortices (see Figure 4). This finding 
can be explained
in terms of the Burger number of the flow 
as follows (see e.g. Polvani et al. 1994). 
Let us define a Rossby deformation radius as 
\begin{equation}  
L_R= \frac{c_s}{2 \Omega}= \frac{H}{2}. 
\end{equation} 
The flow on scales lager than $L_R$ is affected by the Coriolis force
(the Coriolis force becomes of the order of the pressure gradient).
Now, the Burger number is defined by 
\begin{equation}  
B = \left( \frac{L_R}{L} \right) ^2 , 
\end{equation} 
where $L$ is the typical length scale of the flow. 
In a two-dimensional Keplerian disk 
\begin{equation}  
B= \left( \frac{H}{2r} \right) ^2 = \frac{{\cal{M}}^{-2}}{4} ,
\end{equation} 
where ${\cal{M}}$ is the azimuthal Mach number of the flow. 
Therefore, for thin disks
the Burger number becomes very small. However, it has been observed  
(see e.g. Polvani et al. 1994) that for  
small Burger numbers, prograde (anticyclonic in disks) vortices  
are surrounded by rings of adverse (cyclonic) vorticity.  
We found indeed that the  
anticyclonic vortices are shielded by a weak cyclonic vorticity
edge (rather than by a cyclonic vorticity ring). 
The cyclonic vorticity is located (radially) at the outer edge of the
anticyclonic vortex.  
Although cyclonic vorticity perturbations decay in the flow within a few
orbits, the cyclonic edges do not. 

\subsubsection{Accretion of Clumps of Gas} 

In order to explore the idea that the impacts of clumps of gas onto 
a protoplanetary disk can generate vortices, we first simulated a related
problem, in which previous research suggested that vorticity
is created. 
Namely, the idea that the deposition of energy 
due to the impact of a comet onto  
a planetary atmosphere creates a vortex at the impact site. 
Harrington et al. (1994), for example, 
simulated the dynamic response of Jupiter's atmosphere
to the impact of the fragments of 
comet Shoemaker-Levy 9. In all of their simulations, they obtained
both a set of globally-propagating inertia-gravity waves (the speed of 
propagation of the waves was $400 ms^{-1}$) and a longer-lived 
vortex at the impact site. \\  

We first wanted to test whether we can reproduce these results,
using our numerical tools. We therefore carried 
out a similar simulation using a 
two-dimensional Fourier pseudospectral code (the code was written 
and developed in this research specifically for this  
purpose). We modeled the atmosphere 
assuming that it is a two-dimensional polytropic compressible flow. 
In order to simulate the inertia-gravity waves we chose  
the polytropic constant $K$ in such a way that the speed of the waves matched
$400 ms^{-1}$, and the polytropic index $n$ was set to 1 (this is mathematically 
equivalent to solving
the shallow water equations). The models are insensitive to the
precise values of $K$ and $n$, and qualitatively similar results are obtained   
when simulating sonic waves rather than the inertia-gravity
waves (the polytropic constant 
is then chosen in such a way that the propagation 
speed of the waves matches the sound speed of $700 ms^{-1}$). 
At the impact site we deposited $10^{28}erg$, corresponding to a fragment  
1-km in diameter (of density $1 g~cm^{-3}$, 
and mass $5\times 10^{14}$g). \\  

We obtained the same results as in Harrington et al. (1994): 
a set of globally-propagating
waves and a longer-lived vortex at the impact site (Fig. 5).
The vortex was present during the entire simulation, which was carried
out for about 30 Jovian days. We obtained similar results for models
rotating more slowly (representing the Earth; in the case of the Earth
the sonic crossing time is of the order of the rotation period, 
while for Jupiter it is much longer). The mechanism by which vorticity
is created at the impact site is the following. Due to the Coriolis 
force, motion towards the poles is deflected to the East, 
while motion towards the equator is deflected to the West. 
Therefore, the outward motion due to the high pressure gradient 
at the impact site generates an anticyclonic vorticity motion.  \\ \\

Next we examined the possibility that a local deposition of energy in a 
Keplerian disk
can generate vorticity as well. In this case it is mainly the shear, 
rather than the Coriolis force, that would be responsible for 
inducing the anticyclonic motion.  
We assumed that energy can be deposited in the same manner, due to 
the collision of an infalling clump of gas (a remnant 
of the initial collapsing cloud; see e.g. Cassen \& Mossman 1981) 
with the forming disk. For definiteness 
we assumed that the orbit of the clump was similar to that of the 
Shoemaker-Levy 9 comet, and that it impacted the disk at a 
distance of 5AU from the central star. In this case the kinetic
energy of the clump at impact scales linearly with the mass the
fragment. We
carried out simulations with a maximum energy input of $10^{37}$ ergs,
corresponding to a maximum clump mass of $5\times 10^{23}$ gm. 
\\ 

In the models that we ran, we found that initially, an anticyclonic vorticity 
region formed  
at the impact site (Fig. 6a). However, within a short time, the vortex 
was strongly sheared by the flow and no coherent structure was observed. 
Specifically, the anticyclonic vorticity stripe completely dissipated 
within a few orbits (Fig. 6b). 
The reason for the rapid destruction was the fact that the vorticity 
amplitude was too small for the vortex size (or equivalently the  
size was too large for the amplitude). Any vorticity perturbation with
a characteristic velocity disturbance $u$ needs to be much smaller in size than
the characteristic length $L_s$ related to the shear by $L_s = \sqrt{u/
\Omega '} $, otherwise it is quickly destroyed. 
We have simulated impacts with clouds of size up to ~$H$, with a desnity of
$10^{-4}$ of the local density in the disk. Since this corresponds to masses
that are as high as $10^9$ times the mass of a typical Shoemaker-Levy 9 fragment,
we have to conclude  
that this mechanism is unlikely to produce vortices
in protoplanetary disks.  
\\ \\

\subsubsection{Convective Cells} 

A second possibility (in principle) is that vortices are formed by convection.
The idea is that convective bubbles 'rotated'
by the shear could generate vorticity in the flow. 
Since our simulations are two-dimensional (and therefore cannot
follow vertical convection), we investigated ways in which the
effects of convection could be mimicked. 
If we assume convective cells to be 
of size $l=aH$ (where the "mixing length" parameter $a$ is smaller than one),
the difference in velocity across the cell, due to the shear, is 
given by $\Delta v = l \times dv_K/dr \approx  a c_s$. 
We found that vorticity perturbations with velocity differences corresponding
to a value of $a \approx 0.1$ (when $H/r \approx 0.1$) were sufficient 
to create a vortex in the flow. However, to concentrate dust in the core of 
an anticyclonic vortex one needs to have $a \approx 0.2$ (see \S 3.2; 
this could also  
be achieved by mergers of vortices). We therefore simulate the 
formation of vortices due to convection by generating in the flow 
vorticity perturbations of size $l \approx 0.1 H$ and velocity
$v \approx 0.1 c_s$. We introduce the perturbations in the flow 
at the rate of about one per orbit (since $v/l \approx \Omega_K$). 
In this case the simulations are similar to those of driven turbulence.  \\ 

We ran this model for many rotation periods, allowing the disk to
evolve. The simulation showed that with time, the vortices that formed
interacted and merged together. 
Eventually, after about 50 orbital periods, one large vortex dominated 
the flow (Figure 7). \\

\subsection{Dust Dynamics}

The time scale parameter $\tau=1/\gamma$ in eqs.2 and 3 is defined to be
approximately the time it takes a (spherical)
dust particle to basically come to rest in the flow, and it is given
by (e.g. Barge \& Sommeria 1995):
\begin{equation}
\tau = \frac{\rho_d s}{\rho_{gas} c_s}=\frac{2 \rho_d s}{\Sigma \Omega}, 
\end{equation} 
where $\rho_d$ is the density of the particle, $s$ is its radius,
$\rho_{gas}$ is the gas density, $c_s$ is the speed of sound,
$\Sigma $ is the gas surface density and $\Omega$ is the (Keplerian)
angular velocity. \\  
 
For very light particles, the stopping-time $\tau$ is short compared to 
$\Omega^{-1}$ and the particles come
to rest rapidly (relatively to the flow). In this case the dust just moves 
with the flow. For heavy particles, one has $\tau >> 1/ \Omega$, and the dust
particles follow an almost Keplerian motion without being affected by the drag
forces. For intermediate mass particles (for which $\tau \approx 1/\Omega$) 
it is expected that the vortex captures particles in its vicinity
(e.g. Barge \& Sommeria 1995). Following Cuzzi et al. (1993),  we assume
that the mean free path $\lambda$ in the gas is given by 
$\lambda \approx (r/1 AU)^{11/4}$ cm. The equations for the dust particles
(eqs.2-3) are valid in the Eptstein regime, i.e. when the radius $s$ of 
the particles is smaller than the mean free path: 
$s < \lambda$. 
Assuming that $\rho_d=3$g/cm and taking  
$\Sigma = \Sigma_0 r^{-1.5}$ where $\Sigma_0=1700 g/cm^2$ and 
$r$ is given in AUs (Barge and Sommeria 1995), 
the stopping parameter
$\tau $ takes the value $1/\Omega(r)$ at $r_0=60AU$ for $s=1$cm and
at $r_0 \approx 5AU$ for $s=50cm$. Therefore, our simulations are valid
in the range $r > r_0/60$ for $s=1$cm and $r > 0.8 \times r_0$ for $s=50$cm
(where we have defined $r_0$ to be the radius at which $\tau=1/\Omega$ for
particles of a given size). Since most of the simulations were carried out
in the regime $\tau \approx 1/\Omega$, the results are valid only for 
particles of radius $s \approx 50$cm or smaller, otherwise the 
dust particles are not in the Epstein regime at the radius where 
$\tau \approx 1/\Omega$. The results are also valid for
chondrules for which $s \approx 0.1$ cm, where the Epstein regime
is realized for $r>0.4$ AU, and $r_0 \approx 200 AU$. 
All the models scale with radius $r_0$ and orbital period
$P=2 \pi /\Omega (r_0)$.      \\ 

\subsubsection{Concentration of dust particles in the core of anticyclonic
vortices} 

We carried out simulations in which a single vortex was initially 
introduced in the disk, with $H/r=0.15$, and $H/r=0.5$. The second
case, $H/r=0.5$, was intended to mimic a disk similar to that 
of Bracco et al. (1999), where the vortices were not very elongated. 
In both cases the initial velocity 
in the vortex was taken to be
 a significant fraction of the sound speed (more
than 30 percent). The results obtained for both disks are similar. 
The initial random distribution of dust particle
in the disk is shown in Figure 8. 
With time, dust particles concentrate inside the vortex 
(Figures 9 and 10). The number of dust particles in the core of the
vortex increases linearly with time (Figure 11). 
The particle density inside the core was doubled (in comparison to the ambient 
particle density) within about 3 orbits. 
The radial drift of the particles is also clearly visible in the outer
disk (Figure 10).    
\\ 

\subsubsection{Radial drift of the dust particles in the disk} 

Because of the compressibility, the gas flow is partially supported by
pressure forces, and its
angular velocity is slightly sub-Keplerian.  
For dust particles of a given size and density, one has
$\tau \propto r^{3}$ (Eq.7).    
At the radius $r_0$, at which $\tau \approx 1/ \Omega_K$,
the drag exerted by the gas flow on the particles becomes non-negligible
and the particles are slowed down to sub-Keplerian speed. The centrifugal
force is consequently decreased and the particles drift radially inwards. 
At larger radii, however, one has $r>>r_0$  
and $\tau >> 1/\Omega_K$. The drag there is negligible and the particles 
rotate at a Keplerian velocity, without being affected by the gas flow. 
On the other hand, at small radii ($r<<r_0$),  
the drag is so strong ($\tau << 1/\Omega_K$) that 
the particles move completely with the flow at a sub-Keplerian velocity without 
even drifting inwards (i.e. like tracers).  
One might therefore expect
a gap to form around the location where $\tau \approx 1/\Omega_K$,  
with the matter having the tendency to accumulate at a radius $r< r_0$. 
\\  

We found that the radial drift velocity reaches a maximum value of 
about $ v_{drift} \approx 10^{-2} \times v_K$ (when $\tau \approx
1/\Omega_K$).    
Accordingly, the timescale for the particles to drift inwards is
very short ($t_{drift} \approx r/v_{drift}$),  
of the order of $\approx 20$ Keplerian orbits.
We found that the surface density of the dust particles in the inner disk
(i.e. $r<r_0$) increased by a 
factor of 2-3 within about ten rotation periods 
(Figure 12). However, the depletion of dust in the outer disk is slower
($\approx 100$ orbits), 
since a much broader region of the disk is involved.   

\subsection{Transport of Angular Momentum} 

It is interesting to consider whether interactions and 
mergers of vortices can result in fignificant 
angular momentum transport in the disk.  

The effective "viscosity" parameter, $\alpha_{eff}$, is given
in a steady state by (e.g. Pringle 1981)  
\begin{equation} 
\alpha_{eff} = \frac{2}{3} \frac{r}{H} \frac{v_r}{c_s}, 
\end{equation} 
where $v_r$ is the radial velocity.
In Figure 13 we show $\alpha_{eff}$ as a function of time for 
the merger of two vortices.  
It is clear that for such a mechanism to provide efficient
angular momentum transport, 
it requires merging of vortices every few rotations. 
Therefore, one needs a process that is able to generate vortices
continuously (and a low MHD viscosity to avoid their destruction;
Godon \& Livio 1999). The only process capable (at least 
in principle) of such vorticity generation is convection, however,
it would require a full three-dimensional calculation to assess the 
viability of this mechanism. At the moment it appears unlikely that
vortices would play an important role in angular momentum transport.
In particular, even if vortices are formed continuously, 
it appears difficult for mergers to sustain a relatively coherent
outward flux of angular momentum.

\section{Discussion} 

In this work, we have carried out for the first time a 
two-dimensional compressible simulation 
of a Keplerian disk,  with the purpose of studying the {\it formation} 
and {\it role} of vortices in protoplanetary disks.  
We found that in order to generate a vortex in the disk, 
the initial vorticity perturbation has to be anticyclonic and 
relatively strong: 
its velocity and size have to be a considerable 
fraction (at least $\approx 0.1$)
of the sound speed and disk thickness respectively.   
We also found that each anticyclonic vortex is shielded by 
a weak cyclonic vorticity stripe ("vortex shielding"), while
the cyclonic perturbations are elongated and sheared by the
flow.  \\  

We showed that if the disk that forms after the collapse of
the protostellar cloud is initially turbulent and contains a randomly
perturbed vorticity field, then coherent
anticyclonic vortices can form, and they merge together into larger vortices. 
The vortices so formed decay slowly, on a timescale that is inversely proportional
to the viscosity parameter, and of the order of 50-100 orbits for
$\alpha_{SS} \approx 10^{-4}$. \\   

It is important to note that the decay time of the vortices
for disks with parameters ranging from $(H/r,\alpha)=(0.5,10^{-5})$ to 
$(H/r,\alpha)=(0.05,10^{-3})$
(Godon \& Livio 1999) was the same as the one obtained here, because
the viscosity is given by $\nu = \alpha_{SS} c_s H = \alpha_{SS} H^2 \Omega_K$.
We can, therefore, infer the same decay time for a disk with 
$\alpha_{SS}=10^{-2}$ and $H/r=0.015$. 
This is important since the the maximal particle size
reached by coagulation (cm) was calculated using $\alpha_{SS}=10^{-2}$
(more precisely, the size is given by $150 \alpha_{SS}$ cm; Dubrulle, Morfill
\& Sterzik 1995). Such a value of the aspect ratio $H/r=0.015$, 
however, would have
required a higher resolution and was therefore less convenient from
a numerical 
point of view. Furthermore, value for $\alpha_{SS}$ in the range
$10^{-4}-10^{-3}$ are frequently used to model the FU Ori outburst cycles
in young stellar objects (e.g. Bell et al. 1995). \\  

We also attempted to generate vortices by simulating the impacts of 
accreting clumps of gas onto the disk. However, the anticyclonic 
vorticity perturbations that form as a result of such impacts
are not strong enough to generate  
a coherent structure, and they dissipate within several orbits. \\

We have in addition carried out simulations of driven turbulence, to 
mimic the effects of convection, where vorticity is continuously 
generated in the flow. These simplified calculations showed 
that after about 50 orbits
a large vortex forms and is sustained in the flow, due to the merging 
of smaller vortices. Such a vortex is a preferred place for 
the dust to concentrate and trigger the formation of a large planetesimal
or a core of a protoplanet. \\  

In addition to vortex formation, we examined the process of dust 
concentration by carrying out simulations of a two-phase flow 
designed to model the dust-gas interaction in protoplanetary disks.
We found that the dust concentrates quickly in the cores of vortices when 
the drag parameter is of the order of the orbital frequency.
As a consequence of the radial drift, particles are continuously renewed 
near the vortex orbit. The dust density in the vortex increases by a 
factor of 10 within about 20 orbits. \\ 

In the Introduction, we noted  
that there is a gap of two orders of magnitude 
between the maximal particles size (centimeters) reached by coagulation  
and minimal size (meters) required for planetesimal formation 
(via a gravitational instability). 
The minimal size required for planetesimal formation could be reduced 
(to centimeters), if the the density in the vortex
is increased by a factor of ~100. According to our simulations
this would happen in about 200 orbits,
or about 2000 years at 5AU - a short time compared to the timescale of the 
the formation of
centimer-size objects ($10^4$ years; Beckwith, Henning, \& Nakagawa 1999).
We found that vortices survive in the flow for about 50 orbits
(unless the alpha viscosity parameter is smaller, or vortices are
constantly generated in the flow), thus reducing the gap for this
process to work from a factor of a 100 to 4. \\  

We also found that the particles drift rapidly inwards, 
due to the compressibility of the flow. This result was not
obtained in previous two-dimensional incompressible models
of the disk. 
The drift of the dust induces a significant increase in the surface density 
of the dust particles. In about 10 orbits the density can be increased 
in the inner disk 
by a factor of about 3, and in the outer disk it is concomitantly
decreased by about
the same factor within a hundred orbits. 
Specifically, for dust particles of radius 10cm and density 3g/cc,
we find that within about 300 years (about 10 orbits) the density 
increases (by a factor of 3) in the region $r< 9$AU, and 
it decreases at larger radii ($r > 9$AU) in about 3000 years.  \\  

We have also considered the effects of interactions and 
mergers of vortices on angular momentum transport in the disk.  
We found that even if vortices are formed continuously, 
it appears difficult for mergers to transport  
angular momentum outwards effectively.  \\  
 
\section*{Acknowledgments} 
This work has been supported 
by NASA Grant NAG5-6857 and by 
the Director's Discretionary Research Fund at STScI. 
We would like to thank James Cho for useful
discussions on the vorticity equation. \\  


\section*{References}

\noindent       
Adams, F. C., \& Watkins, R. 1995, ApJ, 451, 314 
\\ \\          
Barge, P., \& Sommeria, J. 1995, A \& A, 295, L1 
\\ \\
Beckwith, S. V. W., Henning, T., Nakagawa, Y., 1999, in Protostars
and Planets IV, in press, astro-ph/9902241  
\\ \\   
Bell, K. R., Lin, D. N. C., Hartmann, L. W., \& Kenyon, S. J.,
1995, ApJ, 444, 376 
\\ \\ 
Bracco, A., Chavanis, P.H., Provenzale, A., \& Spiegel, E.A. 1999,
preprint, astro-ph/9810336. 
\\ \\ 
Bracco, A., Provenzale, A., Spiegel, E., Yecko, P. 1998, in A. Abramowicz,
G. Bj\"ornsson, J.E. Pringle (ed.), Theory of Black Hole Accretion Disks,  
Cambridge Univ. Press, 254  
\\ \\          
Cassen, P., \& Moosman, A. 1981, ICARUS, 48, 353 
\\ \\          
Cuzzi, J.N., Dobrovolskis, A.R., Champney, J.M. 1993, ICARUS, 106, 102 
\\ \\ 
Dubrulle, B., Morfill, G., and Sterzik, M. 1995, ICARUS, 114, 237 
\\ \\          
Farge, M., \& Sadourny, R. 1989, J. Fluid Mech., 206, 433 
\\ \\          
Fridman, A.M., \& Khoruzhii, O.V. 1999, in Astrophysical Discs, ASP
Conference Series, Vol. 160, J.A. Sellwood and J. Goodman, eds., 341 
\\ \\ 
Godon, P., \& Livio, M. 1999, ApJ, 523, 350  
\\ \\ 
Godon, P. 1997, ApJ, 480, 329 
\\ \\          
Harrington, J., LeBeau, R.P.Jr, Backes, K.A., \& Dowling, T.E, 1994,
Nature, 368, 525. 
\\ \\ 
Lovelace, R.V.E., Li, H., Colgate, S.A., \& Nelson, A.F. 1999, ApJ, 513, 805
\\ \\ 
Nauta, M.D., 1999, A \& A, in press 
\\ \\       
Polvani, L.M., McWilliams, J.C., Spall, M.A., \& Ford, R., 1994,
Chaos, 4, 177 
\\ \\ 
Pringle, J.E., 1981, ARA\& A, 19, 137 
\\ \\ 
Shakura, N.I., \& Sunyaev, R. A. 1973, A \& A, 24, 337. 
\\ \\       
Tanga, P., Babiano, A., Dubrulle, B., \& Provenzale, A., 1996, ICARUS, 121, 158 
\\ \\       
Tassoul, J. L., 1978, Theory of Rotating Stars, Princeton Series in
Astrophysics, Princeton, New Jersey 
\\ \\       
%

\newpage 

\section*{Figures Captions}

\noindent
{\small{\it Figure 1: Color scale of the initial random perturbation
of the vorticity field. 
Green, yellow and brown represent increasing anticyclonic vorticity,  
and cyclonic vorticity is in dark blue (light blue represents null
vorticity). 
The Keplerian background has been subtracted for clarity. 
}} \\  

\noindent
{\small{\it Figure 2: A color scale of the vorticity is shown at t=8 orbits
for the model shown in Figure 1.
Coherent anticyclonic vortices have formed and merged into larger vortices. 
Cyclonic vorticity appears only as elongated (dark blue) stripes.
}} \\  

\noindent
{\small{\it Figure 3: The spectrum of the total kinetic energy
(averaged in time and in the radial direction) is shown as a function of the 
azimuthal wavenumber for the model shown in Figures 1 and 2. 
The higher
modes are smoothed out by the viscosity and there the slope is $\approx 
-2.8$. In the lower modes the slope is fairly flat ($\approx -0.6$),
as expected for compressible two-dimensional flows.  
}} \\  

\noindent
{\small{\it Figure 4: Vortex shielding in a disk. 
A color scale of the vorticity in the disk is shown.  
The Keplerian background has been subtracted for clarity.
The anticyclonic vortex is accompanied by a cyclonic vorticity stripe 
(dark blue band) stretching azimuthally. 
Another weak
anticyclonic vortex is seen (in green).
}} \\  

\noindent
{\small{\it Figure 5: A simulation of the Jovian atmosphere about 48h after
impact. The color scale represents the vorticity. Waves propagate
outwards (at the inertia-gravity waves speed) 
and a coherent vortex forms at the impact site. 
}} \\  

\noindent
{\small{\it Figure 6a: Colorscale of the vorticity in a disk after energy
has been deposited locally, at $t \approx 0.1 $ orbit. An anticyclonic vortex
forms at the (rotating) impact site, and it stretches with time. A wave
propagates outwards and is strongly deformed by the shear. The Keplerian
background vorticity has not been subtracted for comparison.  
}} \\  

\noindent
{\small{\it Figure 6b: The vorticity of the disk after a little 
more than one orbit.
The vortex has been stretched out to the point that it is now barely visible
in the lower left.   
}} \\  

\noindent
{\small{\it Figure 7: A grayscale of the vorticity in the disk after about 50
orbital periods. The simulation here corresponds to driven turbulence, namely,
an anticyclonic vorticity perturbation is introduced in the flow every orbit.
Such a perturbation is first stretched and then 'collapses' onto itself to form
a coherent vortex. The vortices so formed interact together and merge into
larger vortices. Eventually, as shown here, one large vortex dominates the
flow (in the upper part of the figure). 
The Keplerian background has been subtracted for clarity.  
}} \\  

\noindent
{\small{\it Figure 8: The initial random distribution of dust particles in the
disk. The number of particles is $N=15,000$ and 
the surface density of the particles increases like 1/r.  
}} \\  

\noindent
{\small{\it Figure 9: A strong anticyclonic vortex in the disc, 
shown at t=12 orbits. 
}} \\  

\noindent
{\small{\it Figure 10: The distribution of dust particles in the
disk after 12 orbits. The density of the dust particles in the vortex
has increased by a factor of five in comparison to its initial value.
The drift of the particles inwards is also apparent. More than a half 
of the particles have crossed the inner boundary of the computational 
domain. 
}} \\  

\noindent
{\small{\it Figure 11: The number of particles located inside the vortex as a 
function of time. The number of particles $N(t)$ is in units of $N_0
=N(t=0)$, and the unit of time is the orbital period of the vortex in 
the disk. The dotted line represents a thick disk ($H/r\approx 0.5$)
and the full line is for a thinner disk with $H/r =0.15$. 
}} \\  

\noindent
{\small{\it Figure 12: The dust particles in the disk at 
$t\approx 10$ orbits. The inner radius of the disk is at $r_{in}=1$
and the outer radius is at $r_{out}=41$. The radius at which 
$\tau \approx 1/\Omega$ is located at $r=r_0=10 \times r_{in}$. 
Due to the drift, the particle density in the inner disk has 
increased by a factor of about 3.  
}} \\  

\noindent
{\small{\it Figure 13: The effective alpha viscosity parameter $\alpha_{eff}$
(due to the merger process of vortices) 
in a disk (in which $\alpha_{SS} = 10^{-4}$).  
During the first orbits
the vortices emit waves and $\alpha_{eff}$ reaches its maximum.
Eventually the vortices merge together at around $t\approx 12$ orbits.  
}} \\  

\end{document}